\documentclass[9pt,twocolumn,twoside]{optica}
\setboolean{shortarticle}{false}
\setboolean{minireview}{false}

\usepackage{fp}
 
\newlength{\TOarg} \newlength{\TOunit}
{\catcode`p=12 \catcode`t=12 \gdef\TOnum#1pt{#1}}
\newcommand\TOop[2]{\setlength{\TOarg}{#2}%
    \FPdiv\TOres{\expandafter\TOnum\the\TOarg}{\expandafter\TOnum\the\TOunit}%
    \FPround\TOres\TOres{#1}}

\title{Multi-wavelength arbitrary waveform generation through spectro-temporal unitary transformations}

\author[1,2,$^{\ast}$]{Mikael Mazur}
\author[1,${\dagger}$]{Nicolas K. Fontaine}
\author[1]{Haoshuo Chen}
\author[1]{Roland Ryf}
\author[1]{David T. Neilson}
\author[1]{Gregory~Raybon}
\author[1]{Andrew~Adamiecki}
\author[1]{Steve Corteselli}
\author[2]{Jochen Schr{\"o}der}

\affil[1]{Nokia Bell Labs, Crawford Hill Lab, 791 Holmdel-Keyport Rd, Holmdel, NJ, 07733, USA}
\affil[2]{Photonics Laboratory, Chalmers University of Technology, Gothenburg, SE-41296, Sweden}

\affil[*]{Corresponding authors: mikael.mazur@chalmers.se,  ${\dagger}$nicolas.fontaine@nokia-bell-labs.com} 

\begin{abstract}
Temporal waveform manipulation is a fundamental functionality in optics and crucial for applications like optical communications, microwave photonics and quantum optics. 
Traditional IQ- or phase-amplitude modulators shape light by carving energy from the input lightwave and are thus fundamentally lossy, and cannot apply independent modulation to multiple input wavelengths simultaneously. 
Taking inspiration from the space-time duality, we produce arbitrary unitary spectro-temporal transformations on multiple temporal input vectors with a modulation structure comprising of only lossless phase modulation and dispersive allpass filtering. 
The bandwidth of the output waveforms is not restricted by the driving electronics and independent transformations can be performed simultaneously on multiple orthogonal inputs such as spectrally separated frequency tones. 
This overcomes the main limitations of traditional electro-optic modulators and offers fundamental new insight into temporal wave manipulation. 
\end{abstract}

\setboolean{displaycopyright}{true}

\setboolean{displaycopyright}{False}
\doi{}

\begin{document}

\maketitle

\section{Introduction}
In the most general sense, temporal modulation can be considered as a transformation between an input temporal sequence or time-vector and a desired corresponding output vector. 
To fully control the light field it is necessary to manipulate both the phase and amplitude of the input lightwave. Using IQ-modulators, the modulation process exhibits a fundamental minimum loss of 3\,dB due to the required combiner while in practice losses often exceed 10\,dB~\cite{Fujitsu100GMod}. 
Moreover, multiple spectrally separated input waves cannot be modulated independently without separating and modulating the input waves in parallel using multiple modulators.
The strong desire to increase the available modulation bandwidth has driven research in new integrated material platforms, 
such as Lithium Niobate on insulator (LNOI)~\cite{Wang2018a}, plasmonics~\cite{Haffner2015}, or organics~\cite{Koos2009}, as well as the integration of parallel modulators and wavelength multiplexers to flexibly manipulate inputs with consisting of multiple spectral lines~\cite{Fontaine2010}. While this has allowed to increase the total bandwidth, all these devices still rely on the classical principle for IQ-modulation prohibiting unitary operation. 
\begin{figure*}[ht]
\centering\includegraphics{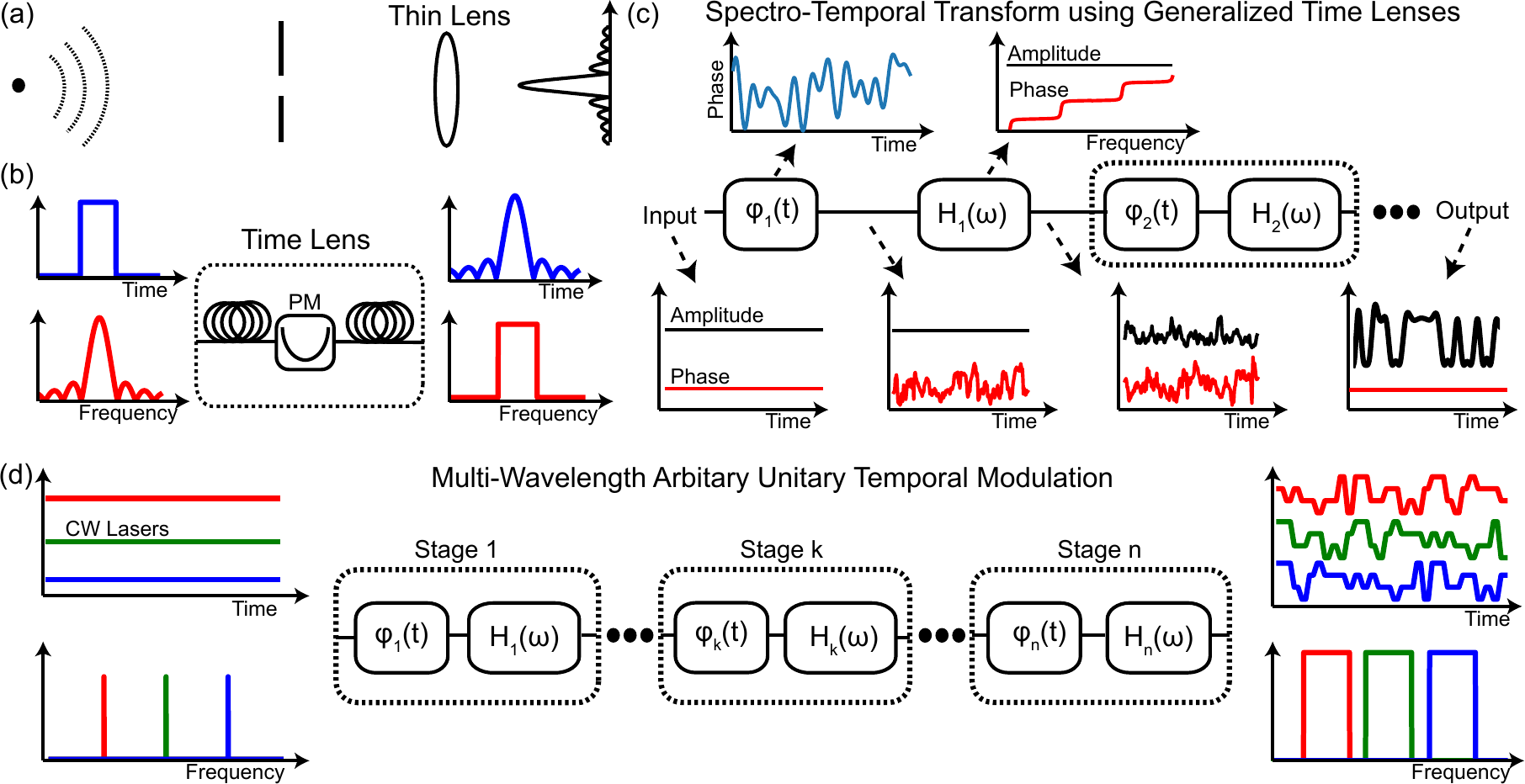}
\caption{\label{fig:timelens}Illustration of the working principle for (a) a standard spatial "thin lens" and (b) an equivalent "time lens" using quadratic dispersion and quadratic phase modulation. 
Although being the most used tool, the "time-lens" implementation only represent a strongly restricted spectro-temporal transform. 
(c) Principle of implementing an arbitrary unitary transformation using generalized time lenses. 
In contrast to the lens-based approach, neither the phase modulation nor the dispersive propagation is restricted to being quadratic. 
Instead, N arbitrary phase modulation, $\phi_k(t)$ and general dispersion $H_k(\omega)$, which can be any all-pass filter shape implementing a dispersive delay, are used to reshape an input CW laser to a target output field. 
(d) Proposed multi-wavelength arbitrary unitary temporal modulation structure consisting of N segments of generalized time lenses. 
By seeding the structure with spectrally separated laser lines and jointly optimize the  phase modulations $\phi_k, k=1,2...n$, independent waveforms can be encoded onto each line. 
This novel application allows for exploiting modulation bandwidths vastly larger than the driving electronics.} 
\end{figure*}

To overcome the bandwidth limitations of traditional modulation, researchers have taken advantage of the broad bandwidth inherent to optics to enhance lightwave manipulation via all-optical signal processing~\cite{minzioni2019}. Many of these experiments have taken advantage of one of the most fascinating aspects of optical physics, the so-called space-time duality~\cite{Kolner1994};
the phenomenon that the propagation of light beams in space governed by paraxial diffraction, and linear propagation of temporal optical pulses in waveguides governed by narrowband dispersion, are described by mathematically equivalent equations. 
The the most basic manifestation of the space-time duality is arguably frequency-to-time conversion via dispersive propagation~\cite{Jannson1981}, the time equivalent of Fraunhofer diffraction. 
However, sophisticated applications typically rely on the so-called time-lens concept~\cite{Kolner1994}, as illustrated in Fig.~\ref{fig:timelens}(a) and (b). Equivalently to the quadratic spatial phase modulation performed by a thin-lens, a temporal lens or time-lens can be constructed by a quadratic phase modulation in time~\cite{Salem2013}. 
Time-lenses have been used for ultra-fast waveform characterization~\cite{foster2008}, time-to-frequency conversion of data signals~\cite{Palushani2012}, temporal magnification and compression~\cite{Salem2009}, temporal cloaking~\cite{Fridman2012,Lukens2013} and ghost imaging~\cite{Ryczkowski2016}. 
While a time-lens "temporal imaging" system can be seen to perform a unitary spectro-temporal transformation, a given lens-based imaging system can only perform a specific transformation with little tunability. 
Moreover, the type of transformations that can be implemented with time-lens systems are limited. 
In particular, a transformation that corresponds to arbitrary temporal reshaping, such as the generation of a long random temporal pattern with a target pulse shape from a CW input, which underpins data modulation, cannot be implemented with such a system. 

Here we show that by extending the concept of space-time duality and taking inspiration from the unitary adiabatic transformation between spatial modes based on multi-plane light conversion (MPLC)~\cite{Morizur2010,Fontaine2019}, it is possible to implement a general unitary temporal transform, that enables completely arbitrary reshaping and waveform generation in the temporal domain.
In contrast to the IQ-modulator, the process is theoretically lossless as the input energy is temporally redistributed through a combination of temporal phase-modulation and dispersion, avoiding any loss from carving and splitting. We expand our initial demonstration of single wavelength unitary temporal manipulation~\cite{Mazur2019b} and show that our proposed method can also simultaneously apply independent modulations to orthogonal temporal input vectors without requiring demultiplexing technology, enabling completely new approaches in multi-wavelength signal generation. 

\section{Arbitary Waveform Generation}
Analogous to MPLC, which is based on a cascade of spatial phase manipulations intertwined with dispersive free-space propagation, our modulator is a cascade of temporal phase modulations and all-pass dispersive filtering as conceptually illustrated in Fig.~\ref{fig:timelens}(c). Each stage can be viewed as a generalized time-lens where both the temporal and the spectral phase modulation can take an arbitrary shape, in contrast to a classical time-lens where the phase modulation is parabolic~\cite{Kolner1994}. 
By cascading multiple stages and jointly optimizing the temporal phase modulations (the spectral modulations are fixed by the dispersive elements), an adiabatic transition corresponding to a unitary spectro-temporal transformation can be realized. 
The principle behind this operation is that the spectral components generated by the phase modulation of one stage are temporally reorganized into amplitude (and phase) variations by the dispersive all-pass filter. In contrast to the IQ-modulator, the input energy is temporally redistributed and the method is thus in theoretically lossless because it avoids any carving loss and splitting losses. 

Since each spectral component experiences a unique delay 
it is possible, for example, to simultaneously apply independent modulations to orthogonal temporal input vectors without requiring demultiplexing technology. This feat, as illustrated in Fig.~\ref{fig:timelens}(d), is unfeasible with traditional IQ-modulators. Furthermore, it directly contradicts the assumption that independent manipulation of lines from a multi-wavelength light source, such as a frequency comb, requires wavelength separation and independent modulation of the lines in a parallel structure followed by wavelength combining. This provides new essential insights to broadband waveform generation and manipulation and in our demonstration, we choose to generate signals which encode fully random information onto an input CW laser, representing the most challenging case of arbitrary waveform generation or temporal reshaping. We note that our technique is a generalization of~\cite{Thiel2017}, which was proposed numerically for single waveform manipulation in the context of quantum optics. 

In order to generate a target waveform requires designing the phase modulation for the specific transformation. The design principle is based on "inverse design"/adjoint optimization techniques. Important to note is that there are multiple ways to solve this multi-dimensional adjoint optimization problem and various algorithms can been used to solve similar problems~\cite{Kirkpatrick1983}. Moreover, while co-design of spectral and temporal phase modulations is possible, the same unitary transformations can be achieved by assuming a fixed, known dispersive filter response separating each stage. Similar to~\cite{Fontaine2017}, we modified the wavefront matching algorithm~\cite{Sakamaki2007} to design the phase masks by finding an approximate solution to the optimization problem using $N$ stages in order to realize the unitary transformation.  
The algorithm can be intuitively understood as seeking a discrete approximation to a smooth transition from the input to the output wave. As a unitary system is, by definition, bi-directional, the algorithm propagates the input field forward and the output field backwards through the structure.  At each stage $k$, the overlap (or error) between the two waves can be calculated as
\begin{equation}
\label{eq:1}
    o_{k}(t) = f(k\Delta z,t)b(k\Delta z,t)^*\exp(j\Phi_k(t)),
\end{equation}
with $k$ denoting the k:th stage, $\Delta z$ the propagation distance between each phase modulation, $t$ denoting time, $f,b$ the forward/backward propagating wave and $\Phi_k(t)$ the phase modulation applied at stage $k$. The total error is then calculated by summing the waveform over all time instances according to 
\begin{equation}
    O_k = \int o_{k}(t) dt.
\end{equation}
To minimize the total error, $\Phi_k(t)$ is updated by a gradient descent algorithm to optimize the phase modulation applied in each stage. The phase error $\Delta \Phi_k(t)$ for stage $k$ is calculated according to 
\begin{equation}
    \Delta \Phi_k(t) = - \text{arg} \left[o_k(t) \exp \left(-j \phi_k(t)   \right) \right],
\end{equation}
with $\phi_k = \mathbf{E}\left(\text{arg} (o_k(t))\right)$ and $\mathbf{E}$ denoting the expectation value. In addition, before updating $\Phi_k$ additional constraints such as a bandwidth-limited phase-modulation or signal quantization to account for digital-to-analog converter (DAC) resolution can be added to maximize the output signal quality. The optimization usually converges to a target solution in about 500-1000 iterations. 

For the multi-channel operation, Eq.~\ref{eq:1} is expanded to sum the error for all wavelength-separated fields according to
\begin{equation}
\label{eq:2}
    o_{kij}(t) = \sum_k f_{i}(k\Delta z,t)b_{j}(k\Delta z,t)^*\exp(j\Phi_k(t)),
\end{equation}
with index $i,j\in[1,2,...,n]$ representing the indices of the $n$  spectrally separated input/output channels. The total error can then be represented in an $n\times n$ matrix with $n$ entries given by $O_{kij} = \int o_{kij}(t) dt$. Following this, the phase modulations are updated using the same approach as for the single channel case with the goal of maximizing $\text{Tr}\left(O_{kij}\right)$. The matrix characterize the transform and for a completely unitary transform it is diagonal. 

\section{Experimental Setup}
\begin{figure}[!h]
\centering\includegraphics{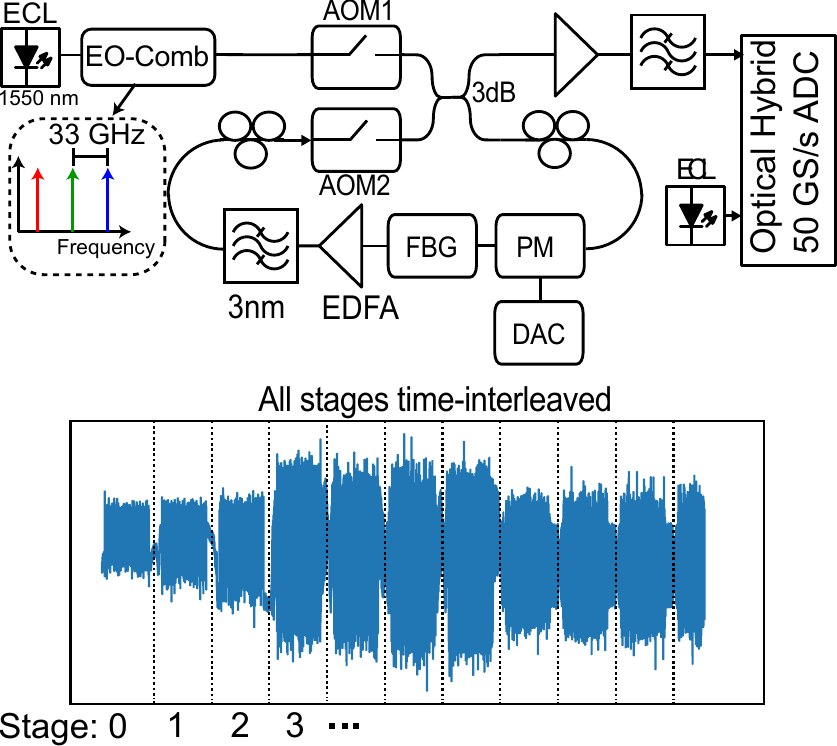}
\caption{\label{fig:setup}Detailed experimental setup used to evaluate the proposed modulation concept. The proof-of-principle demonstration was done using a recirculating loop setup to allow for flexible tuning of the number of stages. The loop was controlled using two fast-switching AOMs. An FBG and a PM inside the loop provided the spectral and temporal phase modulations. An EDFA and a 3\,nm wide optical bandpass filter compensated for loop losses. The signal was coherently detected with an optical hybrid followed by analog-to-digital conversion with a 50\,GS/s real-time oscilloscope. In the case of multi-channel modulation, the channels were generated from the lines of electro-optic (EO) comb and detection was achieved with a free-running ECL  as local oscillator. In the single channel case, a single ECL was split to serve both as modulator input and local oscillator in a homodyne configuration.} 
\end{figure}

The experimental setup is shown in Fig.~\ref{fig:setup}. To allow to flexibly change the number of stages and to capture the output of all stages in a single trace, a recirculating loop setup was used for the evaluation. The loop was controlled using two fast-switching acousto-optical modulators (AOMs) connected via a 3\,dB coupler. The loop input AOM had a rise time of about 5\,ns and was operated using a 350\,MHz RF carrier frequency. Similarly, the AOM inside the loop had a rise time of 10\,ns using a 200\,MHz RF driving signal. As a result, the measured signal was frequency shifted by  $\Delta f = 350 + n\cdot 200$\,MHz with $n$ being the stage index. Inside the loop, the temporal and spectral phase modulations were realized with a standard 30\,GHz bandwidth phase modulator (PM) and a conventional fiber Bragg grating (FBG), respectively. The PM was driven by a 6-bit DAC with about 23\,GHz analog bandwidth. The DAC was controlled using a field programmable gate array (FPGA) with about 10\,Msamples memory. The $V_{\pi}$ of the PM was about 3.5\,V and the DAC output was amplified to about 20\,dBm, resulting in about 0.8\,$V_{\pi}$ of available swing. The FBG was a standard dispersion compensation module with -1321\,ps/nm dispersion covering the full C-band. In addition, polarization controllers before the in-loop AOM and before the PM ensured the state-of-polarization into the PM was aligned for each roundtrip.  

In the single channel evaluation, a narrow-band external cavity laser (ECL) was split to serve as both  modulator input and local oscillator in a homodyne configuration. For the multi-channel configuration, a 33\,GHz electro-optic frequency comb built using a single phase modulator provided the input, as shown in Fig.~\ref{fig:setup}. Given the limited bandwidth of the real-time oscilloscope, it was not possible to detect all channel simultaneously and we therefore used a tunable ECL as local oscillator and measured the three generated waveforms sequentially. 
Although our concept is in principle a loss-less operation, the discrete components in this proof-of-principle experiment had a fairly high insertion losses of about 3\,dB for the PM and 4\,dB for the FBG. In addition, the 3\,dB coupler and the polarization controllers inside the loop added an additional 4\,dB loss. To compensate an Erbium-doped fiber amplifier (EDFA) cascaded with a 3\,nm wide optical bandpass filter to suppress out out-of-band noise was added to the loop. In addition to SNR degradation from the amplification, per-roundtrip polarization fluctuations caused modulation efficiency variations which resulted in a roundtrip-dependent SNR degradation. The resulting trade-off meant that the maximum SNR was limited. The loop output was connected via a second EDFA to an optical hybrid. The output electrical signals were sampled using a 16\,GHz 50\,Gsamples/s real-time oscilloscope. 

\subsection{Data Processing}
For the single channel evaluation, the post-processing consisted of fixed frequency offset compensation and a slow phase aligner which applied a constant rotation over blocks of 2000 samples. The frequency offset estimation compensated the frequency shift induced by the AOMs used to control the recirculating loop. The slow phase tracker compensated fixed phase offsets between the measured trace and the reference signal. These effects were slow varying and originated from mechanical and thermal drifts. No waveform equalization filters were used to improve waveform quality.

For the multi-channel demonstration, an additional frequency offset estimation stage as well as a 4:th power based carrier phase estimation algorithm~\cite{Ip2007} were necessary since homodyne configuration was no longer possible and a residual frequency offset was therefore present after coherent detection. The frequency offset estimation was done by fitting a linear curve to the phase evolution of the received signal. Similarly to the single channel case, no waveform equalization filters were used to improve waveform quality.

\section{Results}
\begin{figure}[t]
\centering\includegraphics{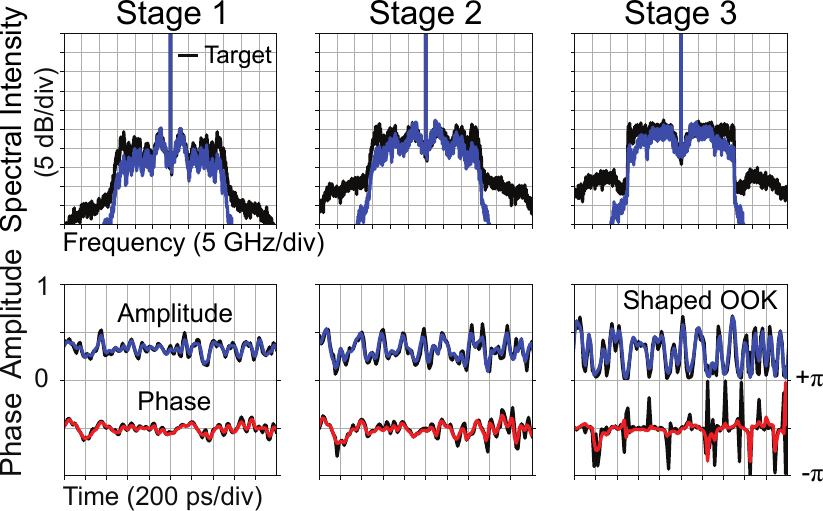}
\caption{\label{fig:single_ook} Generated 20\,Gbaud intensity modulated waveform using 3 stages. The waveform is shaped with a 1\% root raised cosine filter. A correlation coefficient of 88\% was measured between the design target and measured waveforms. } 
\end{figure}
We demonstrate the capability of the proposed modulator with three proof-of-principle experiments, all demonstrating previously impossible functionality.
In the first two experiments we show that our multi-stage spectro-temporal waveform reshaper can perform similar operations to a conventional IQ-modulator by generating chirp-free intensity modulated (on-off keying, OOK) and narrow-band, pulse-shaped quadrature-phase-shift-keying (QPSK) signals encoding random data. Finally we show that our device can simultaneously encode different information on multiple separate wavelength carriers by reshaping three input lines from a frequency comb to independent random output pulse-shaped QPSK waveforms.

\begin{figure}[t]
\centering\includegraphics{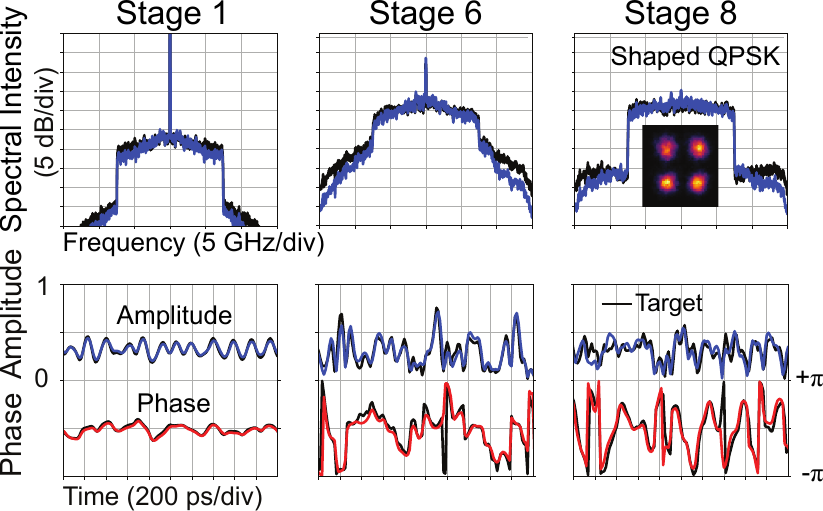}
\caption{\label{fig:single_ch_qpsk} Generated 15\,Gbaud 1\% roll-off QPSK waveform shaped using 8 stages. The output correlation between the design target and measured waveforms was 86\%.  } 
\end{figure}

First, we generated a 20\,Gbaud intensity modulated OOK signal from a CW laser centered at 1550\,nm as shown in Fig.~\ref{fig:single_ook}b. 
The spectro-temporal transform in this case consisted of only three stages.
The spectrum and temporal amplitude and phase of the waveform after each stage is shown in Fig.~\ref{fig:setup}b. 
The output signal contains pure intensity modulation, a rectangular spectrum, and 
the calculated correlation between simulated and measured waveforms is about $88\%$, which is excellent considering the experimental uncertainties.
\begin{figure*}[t]
\centering\includegraphics{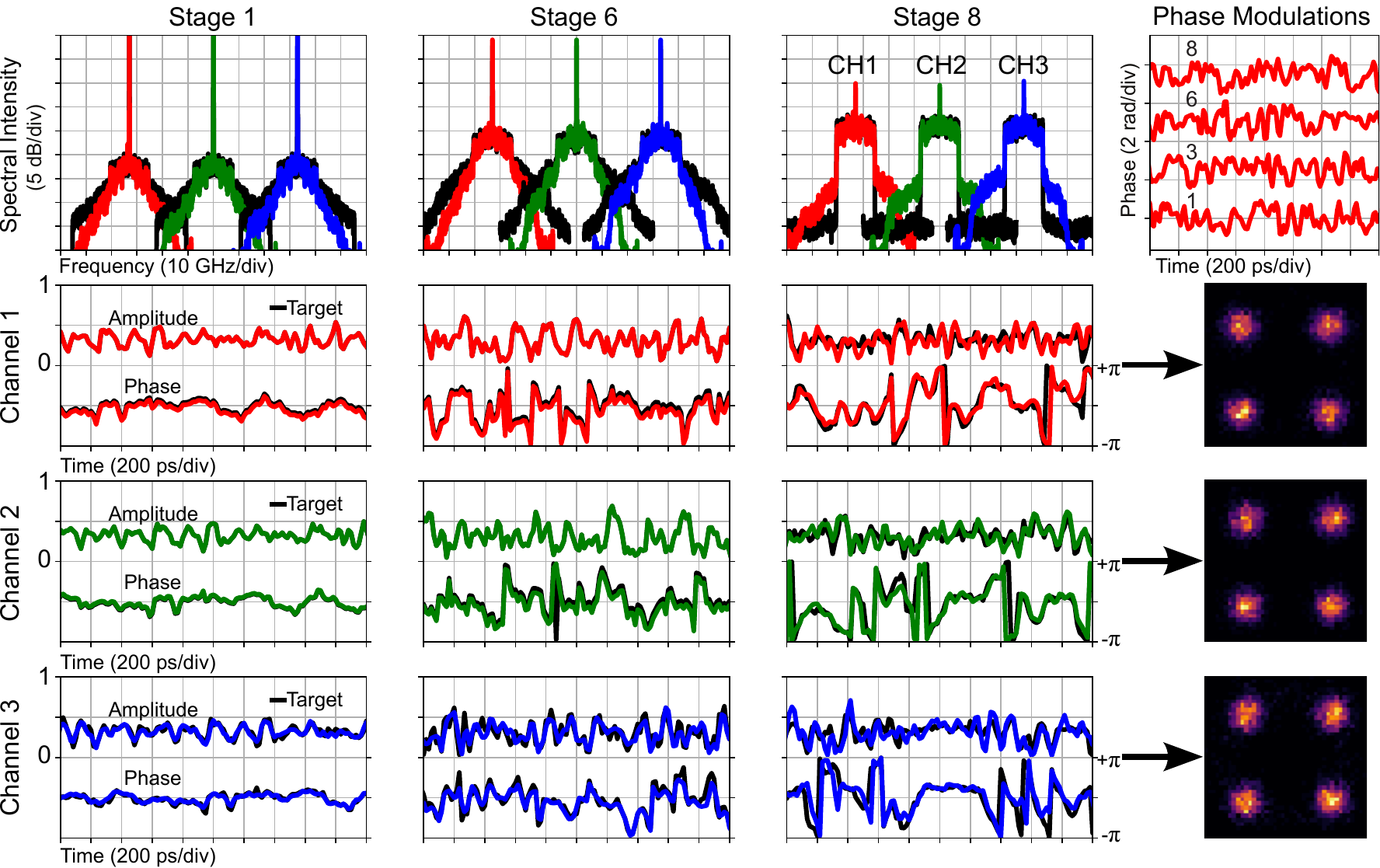}
\caption{\label{fig:multi_ch}Superchannel modulation. The three input laser lines originated from a 33\,GHz-spaced frequency comb and independent, random-data, waveforms were encoded onto each line. The waveforms were spectrally shaped using a 1\% roll-off raised cosine filter.  Similar to the single channel case, eight stages were used to implement the spectro-temporal transform. } 
\end{figure*}

To demonstrate the capability of our device to generate modern coherent communication signals, we generated a 15\,Gbaud QPSK signal, as shown in Fig.~\ref{fig:single_ch_qpsk}. 
Eight stages were used to perform the temporal reshaping. We observe an excellent agreement and a correlation of 86\% between the measured and target waveforms, particularly complete suppression of the DC carrier, indicating that all the energy is successfully redistributed to form the target signal.  Moreover, 
in contrast to the broadened spectrum of a highly QPSK signal generated with traditional phase-only modulation, our signal is strongly bandwidth limited due to the pulse-shaping. 
A density plot of the output constellation is shown in  Fig.~\ref{fig:single_ch_qpsk}, inset. 

The results convincingly demonstrate the capabilities of our device to perform the functionality of an IQ modulator - amplitude and phase modulation with pulse-shaping (i.e. a rectangular spectrum). However, there are numerous fascinating aspects and challenges associated with this modulator. One such aspect is the that increasing the modulation rate by a factor of 2 reduces the dispersion required by a factor of 4. Similarly, because our technique relies on energy redistribution, generating long sequences of zeros would require a large number of stages but such patterns are usually undesired and communication systems employ special coding to avoid them. 
\begin{figure}[!h]
\centering\includegraphics{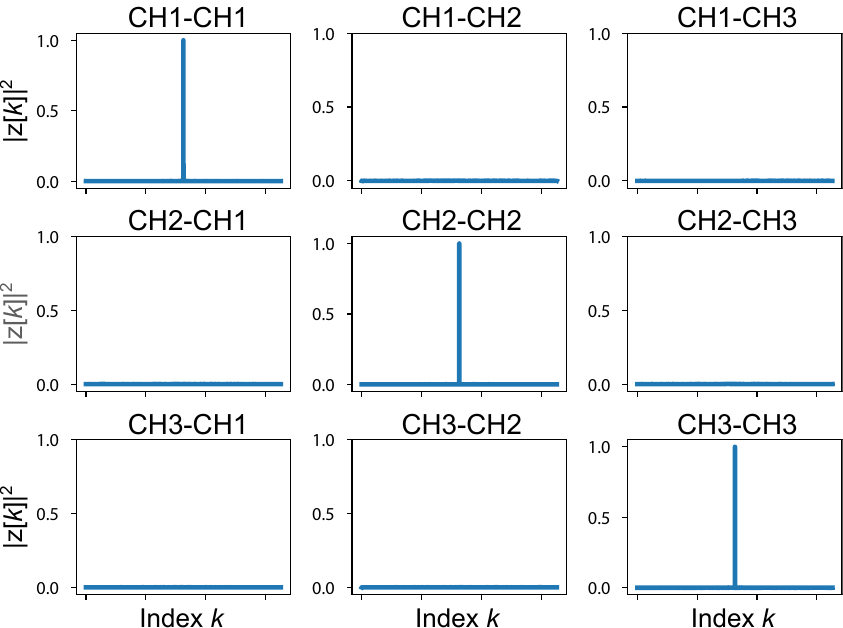}
\caption{\label{fig:sup_wavcorr}Measured cross-correlation between the three output channels. The results diagonal verifies a high quality transform and no coupling between channels is observed. }
\end{figure}

In the final demonstration we modulated three lines from a 33\,GHz spaced optical frequency comb with independent data simultaneously. Note that a comb is not essential and independent lasers on a corresponding grid could also be used. The driving signal to the phase modulator was bandwidth limited to about 20\,GHz, as for the single channel case. Still, the output signal spans a total bandwidth of 90\,GHz, far beyond the capability of a conventional 20\,GHz IQ-modulator and the bandwidth of our electronics. Similar to the single channel case, target waveforms were 15\,Gbaud RRC-shaped QPSK and we used 8 transformation stages. The signals are designed such that each wavelength channel is modulated with an independent random data-pattern at the output. The temporal phase-masks are found by expanding the waveform matching criterion to jointly minimize the error for all three output channels. 

We observe good agreement between the measured and simulated waveforms in
Fig.~\ref{fig:multi_ch} with a correlation of about 85\% for all three channels.
The output constellation diagrams also verify that each channel successfully
generated a high quality waveform. The performance of the
individual channels was comparable to single channel modulation, despite using
only  8 stages to generate the broadband output waveforms. We do observe a
larger residual DC tone in the output spectrum in Fig.~\ref{fig:multi_ch}
compared to Fig.~\ref{fig:single_ch_qpsk}, which we attribute to polarization
mode dispersion (PMD) in the experimental setup.
Finally, to verify the independence of the output waves in the multi-channel experiment, we calculated the cross-correlation between the measured output channels.  
The results are shown in Fig.~\ref{fig:sup_wavcorr}, verifying
the the mutual independence between the measured outputs. 

\section{Summary}
These three demonstrations show how our proposed method is capable of reproducing the functionality of traditional modulators, without their inherent losses. Moreover it is able to generate arbitrary waveforms with bandwidths greatly exceeding the modulator electronics, which allows to simultaneously modulate multiple spectral inputs with different signals, a feat which so for has been impossible without spectral demultiplexing.
Implementing our method in an integrated photonic circuit 
could enable direct application of the concept in various scientific contexts without the specific limitations of our setup. In combination with recent breakthroughs in integrated frequency comb generators~\cite{Herr2013} and LNOI modulators~\cite{Ren2019}, our technique could be exploited to create a micro-scale device with both the multi-channel light source and the temporal reshaper on the same chip. Previous demonstrations in optical communication~\cite{marin2017microresonator}, waveform manipulation in microwave photonics~\cite{Marpaung2019}, broadband waveform generating using waveform stitching~\cite{Fontaine2007} and multi-qubit quantum network with spectral modes~\cite{Lukens2016} all relied on wavelength demultiplexers (often off-chip) and parallel modulators. 
Exploiting the proposed method could significantly reduce losses, complexity and footprint which would greatly accelerate the quest for an highly compact broad-band arbitrary waveform generator which has been the focus of intense research in the last decade. 
We therefore believe the technique has wide ranging applications in optical fields where temporal manipulation of light is desired, including communication and quantum optics. 
As the technique is based on the fundamental properties of waves, its use is not restricted to optics and could find interesting applications in other fields which rely on temporal manipulation of waves, like wireless communication or acoustics.

\section*{Funding Information}
J. S. likes to acknowledge funding from the Swedish research council (VR 2017-05157). 

\section*{Acknowledgments}

The authors would like to thank Magnus Karlsson and Peter A. Andrekson for fruitful discussions.

\section*{Disclosures}

The authors declare no conflicts of interest.



\bibliography{Mazur_References_Cleaned}

\bibliographyfullrefs{Mazur_References_Cleaned}


\end{document}